\documentstyle[preprint,revtex]{aps}
\begin{document}
\draft
\preprint{to appear in Phys. Rev. C. Rapid Communication}
\begin{title}
Microscopic approach to pion-nucleus dynamics
\end{title}
\author{ C. M. Chen and D. J. Ernst}
\begin{instit}
Center for Theoretical Physics and Physics Department,\\ Texas A\&M
University, College Station, TX 77845
\end{instit}
\author{Mikkel B.\ Johnson}
\begin{instit}
Los Alamos National Laboratory, Los Alamos, NM\ \ 87545
\end{instit}
\begin{abstract}
Elastic scattering of pions from finite nuclei is investigated
utilizing a contemporary, momentum--space first--order optical
potential combined with microscopic estimates of second--order
corrections.  The calculation of the first--order potential
includes:\ \ (1)~full Fermi--averaging integration including
both the delta propagation and the intrinsic nonlocalities in the
$\pi$-$N$ amplitude, (2)~fully covariant kinematics, (3)~use of
invariant amplitudes which do not contain kinematic singularities,
and (4)~a finite--range off--shell pion--nucleon model which
contains the nucleon pole term.  The effect of the delta--nucleus
interaction is included via the mean spectral--energy approximation.
It is demonstrated that this produces a convergent perturbation
theory in which the Pauli corrections (here treated as a
second--order term) cancel remarkably against the pion true
absorption terms.  Parameter--free results, including the
delta--nucleus shell--model potential, Pauli corrections, pion
true absorption, and short--range correlations are presented.
\end{abstract}
\narrowtext
Pion--scattering measurements, in combination with phenomenological
descriptions \cite{sici86} of the
propagation of the pion and the delta in the nuclear medium,
have proved useful for probing details of nuclear structure.  The
situation
is not, however, entirely satisfactory because some of the
parameters in these phenomenological descriptions have not been derived
quantitatively, even though their physical origin is believed to be
understood.  Chief among these is a shift \cite{cot80} in the
energy of the two--body, pion--nucleon scattering amplitude, which
is evaluated somewhat arbitrarily.  Therefore, in this work we want to see how
far we can go in
describing the dynamics of the pion and the delta starting from
a purely microscopic
approach in which the dynamics (including the energy at which the in--medium
two--body amplitude is to be evaluated) are completely
determined from theory.  Such an understanding is needed before one can
envision making a reliable extension of the theory
to higher nuclear densities and high temperatures, where the propagation of
the pion in the nuclear medium plays an important role in both heavy--ion
reactions and in astrophysical problems.

{}From the results of such a microscopic
approach we hope to learn the extent to which the existing
phenomenologies are in quantitative agreement with the dynamics as
understood in a variety of contexts,
including what is known
about the reactive content of the interaction (true
absorption, quasi--elastic scattering, and correlation effects),
delta--nucleus dynamics (the delta--nucleus interaction, delta
propagation, and the Pauli principle), and the interplay of the
reaction dynamics with nuclear--structure effects.  Although some
calculations \cite{chiang85,johnson85,sici90,oset87,hirata77} of pion
scattering do
include higher--order terms coming from these effects, a modern, microscopic
test of pion--nucleus
dynamics that makes contact with all this information does not yet exist.

Such a test of pion--nucleus dynamics must deal carefully
with several well--appreciated but technically awkward aspects of the
dynamics.  One is Fermi averaging, which is expressed
as a complicated three--dimensional integral of the off--shell pion--nucleon
scattering amplitude over the nuclear density matrix.  This
integration, when performed without any
approximations, incorporates exactly both the propagation of the delta and
the intrinsic nonlocalities that
are inherent to a two--body resonating amplitude.  Another
is the Lorentz--covariant kinematics.
Finally, the pion--nucleon amplitude utilized should contain explicitly the
nucleon pole.  The neglect of this singularity in the two--body
amplitude leads \cite{ernst78}
to an artificially low momentum cutoff that produces a geometrical
change in the effective radius \cite{ernst79} of the nucleus.  We here make
a test of pion--nucleus dynamics within the framework of the optical
potential which incorporates all of these features.

The isobar--hole model \cite{hirata77}, which
was a successful semi--microscopic approach to the
dynamics, has served as a
phenomenological tool to fit various pion-- (and photon--) induced reactions,
including the true--absorption and quasi--elastic channels.  Much has been
learned about pion and delta dynamics from this model. Even
more has been learned from the abundance of
high--precision data that have been taken at the meson factories
during the ten years since the inception of the model.
Our work relies on this progress to generate a
parameter--free microscopic theory, which
we will compare here to elastic--scattering data from 80 to 226 MeV.

The improvements that we feel to be needed are naturally incorporated by
working in momentum space,
where the various amplitudes can be written in relatively simple analytic
form.  One technical advance which is particularly suited to momentum space
is \cite{gieb85} the use of
``relativistic, three--body, recoupling coefficients.''  These incorporate
exactly Lorentz covariant kinematics \cite{ernst80,gieb82} (including Wigner
spin precession), and they provide natural variables
for performing the Fermi--averaging integral.  The first--order optical
potential
is given in terms of them by
\begin{eqnarray}
\left(\vec{k}_{\pi}^{\prime}\vec{k}_{A}^{\prime}\vert
U\vert\vec{k}_{\pi}\vec{k}_{A}\right)
&&=\sum_{\alpha}\int
{d^{3}k'_n\over 2E'_n}\,{d^{3}k_{A-1}\over 2E_{A-1}}\,{d^3k_n\over 2E_n}
\left\langle
\psi_{k_{A}}^{\alpha}\vert\vec{k}_{n}^{\prime}\vec{k}_{A-1}^{\prime}
\right\rangle\nonumber\\
 &&~~~~~~~~\times\left\langle\vec{k}_{\pi}^{\prime}
\vec{k}_{n}^{\prime}\left\vert t(W_{\alpha})\right\vert\vec{k}_{\pi}
\vec{k}_{n}\right\rangle\left\langle\vec{k}_{n}\vec{k}_{A-1}\vert
\psi_{k_{A}}^{\alpha}\right\rangle\rlap{~~.}
\label{eq:op}
\end{eqnarray}
Here, $\langle\vec{k}_{n}\vec{k}_{A-1}\vert\psi_{k_{A}}^{\alpha}
\rangle$ is the target wave function (labeled by its eigenvalues
$\alpha$), which is proportional to a momentum--conserving delta
function and is a function of the relative momentum of the struck
nucleon and the momentum of the $A-1$ remaining core nucleons.
The pion--nucleon amplitude $\langle\vec{k}_{\pi}^{\prime}
\vec{k}_{n}^{\prime}\vert t(W)\vert\vec{k}_{\pi}\vec{k}_{n}\rangle$
also contains a momentum--conserving delta function and is a function
of the relative momentum between the pion and nucleon.
The three implicit momentum--conserving delta functions that are a part
of the matrix elements on the right of Eq. \ref{eq:op} eliminate all but one
three--dimensional integration (the Fermi--averaging integration, which must
be performed numerically); one overall delta function is left over, and this
conserves the total momentum . In our work,
$t(W)$ is the free--space pion--nucleon T--matrix, but the energies
that appear in it are shifted by an amount that is calculated by
a well--defined prescription designed to minimize the effect of the
higher--order terms in the optical potential.  The
kinematics involved in Eq. \ref{eq:op} are those of a relativistic
three--body
problem with momenta $\vec{k}_{\pi}$, $\vec{k}_{n}$, and $\vec{k}_{A-1}$;
the details of how relativistic recoupling coefficients allow one
to calculate Eq. \ref{eq:op} can be found in Ref. \cite{gieb88}.  We work
with invariant amplitudes \cite{gieb88,ernst90} that are free of kinematic
singularities and utilize invariantly--normed wave functions; these
introduce phase--space factors into the calculation, which are also treated
exactly by working in momentum space.

{}From our discussion, it is evident that in a momentum--space approach
the lowest--order optical potential as we formulate it is quite general and
can be evaluated without approximation.  In this sense, our work
improves not only the phenomenological optical model \cite{sici86} but also
on
numerous aspects of the isobar--hole model \cite{hirata77}, which were both
expressed in coordinate space,
where nonlocalities are not as easily handled.  The propagation of the delta
was fully incorporated in the isobar--hole model, but the integration over the
nonlocalities
associated with the two--body amplitude were approximated by factorization--an
approximation that necessitates \cite{ernst83} a nonnegligible correction,
particularly
for lighter nuclei.  To deal with Lorentz--covariant kinematics, expansions and
further factorizations of integrals were made.  Additionally, the pole in
the two--body amplitude was neglected, a choice which we have been particularly
careful to avoid in order to eliminate the possibility of a spurious
geometrical change in the effective radius \cite{ernst79} of the nucleus.

Given that we are able to calculate the first--order optical potential
without approximation, there remains the question of how to organize
many--body theory (in particular, choosing the energies of the nucleon
and the delta in the medium) to optimize its rate of convergence.  The role
of the energy $W_{\alpha}$ in Eq. \ref{eq:op} is quite important in this
regard
because the half--width of the delta resonance (55~MeV) is the same size
as typical energies that characterize nuclei.  Thus, the results of
a calculation
will be very sensitive to how the energies that constitute $W_{\alpha}$
are chosen.  $W_{\alpha}$ is defined
covariantly as the energy available
in the center--of--momentum frame of the pion--nucleon system,
\begin{equation}
W_{\alpha}^{2}=W_{\pi n}^{2}-\left(\vec{k}_{\pi}+\vec{k}_{n}
\right)^{2}\rlap{~~,}
\label{eq:wcm}
\end{equation}
with $W_{\pi n}$ defined as the energy available to the $\pi N$
pair in the pion--target center--of--momentum frame,
\begin{equation}
W_{\pi n}=W_{0
}-\sqrt{\left(\vec{k}_{\pi}+\vec{k}_{n}\right)^{2}
+m_{A-1}^{2}}\rlap{~~,}
\label{eq:wpn}
\end{equation}
and $W_{0}^{\,2}=S$, the invariant square energy of the reaction.  The
mass of the $A-1$ system, $m_{A-1}$, differs from the mass of the
$A$--body target, $m_{A}$, by a nucleon mass and a binding energy,
$m_{A}=m_{A-1}+m_{n}+E_{b}$.  In their nonrelativistic limit, Eq.
\ref{eq:wpn} is known \cite{liu82} as the ``three--body
energy denominator.''

Utilizing the definition of $W_\alpha$ given in Eq. \ref{eq:wcm} produces
needlessly large higher--order corrections \cite{liu82,ernst85} in the
many--body expansion.  This is
because the delta--nucleus shell--model potential, $U_{\Delta}$,
which is generally believed to be nearly equal to
the potential energy of a nucleon in the nucleus, has not yet been included.
Including the effects of $U_{\Delta}$ in the T--matrix causes an effective
downward shift in the
position of the resonance that tends to cancel the upward shift
caused by the nucleon binding energy in Eq. \ref{eq:wpn}.  To incorporate
this effect, we have
proposed \cite{ernst85} a treatment of $W_{\alpha}$ in Eq. \ref{eq:wcm} that
includes the $U_{\Delta}$ in a first approximation via an energy--dependent
and target--dependent
energy shift.  This shift, called the mean
spectral energy, $E_{ms}$, is derived in Ref. \cite{ernst85} and may be
calculated by
\begin{equation}
E_{ms}(W_{0},)={\int d^{3}r\,\phi_{\pi}^{(-)*}(r)\,
\phi_{\pi}^{(+)}(r)\,\rho(r)\,U_{\Delta}(r)\over\int d^{3}r
\,\phi_{\pi}^{(-)*}(r)\,\phi_{\pi}^{(+)}(r)\,\rho(r)}
\rlap{~~,}
\label{eq:ms}
\end{equation}
where $U_{\Delta}(r)$ is taken to be equal to the shell--model
potential of a nucleon.  In Fig. \ref{fig1} we present results for $\pi^{+}$
elastic scattering
from $^{12}$C at 80, 100, 148, 162, and 226~MeV. The data are from
Ref. \cite{harvey84}  The dashed line is the result of using the full
lowest--order
optical potential, including $E_{ms}$.  The effects including this shift
together with the
binding of the struck nucleons, is quite substantial \cite{ernst85}.  At all
energies shown here
we find that the inclusion of $U_{\Delta}$ is not only
significant but moves the results remarkably close to the data.

At this point, the agreement of the theoretical results with the
experimental data is surprising, because there remains much that
has not been considered.  We know, for example, that the pion true--absorption
channel is about one--half \cite{ashery81} of the total reaction cross
section. The Pauli principle \cite{chiang85,johnson85,dover73} also should
play a significant
role in the scattering of the light--mass pion from the heavier nucleon.
The $p$--wave character of the pion--nucleon interaction produces
nonnegligible correlation corrections which enter in the form \cite{eric66}
of the Ericson--Ericson--Lorentz--Lorenz correction.  We will next
include each of these higher--order terms.  The results will provide
a test of our understanding of each piece of the physics and the
role that it plays in pion--nucleus dynamics.

In order to utilize existing calculations of the second--order terms,
we will make extensive use of the local density approximation.  For
the Pauli and true--absorption terms, we utilize the functional form
of the second--order corrections as derived in Ref. \cite{sici86},
\begin{equation}
U^{(2)}(\vec{k}',\vec k\,)=\lambda_{0}^{(2)}\,\vec k\cdot\vec k'\,
\rho^{(2)} (\vec k-\vec k')\rlap{~~,}
\label{eq:nd}
\end{equation}
where $\rho^{(2)}$ is the Fourier transform of the square of the
target density.
Microscopic calculations of higher--order terms yield a coefficient
$\lambda_{0}^{(2)}$ which itself depends weakly on $r$.  In the
same spirit as the mean spectral energy calculation, we may define
the radius $R_{2}$ at which the pion interacts in a finite nucleus
by
\begin{equation}
R_{2}={\int d^{3}r\,\phi_{\pi}^{(-)*}(r)\,\phi_{\pi}(r)\,\rho(r)\,r
\over\int d^{3}r\,\phi_{\pi}^{(-)*}(r)\,\phi_{\pi}(r)\,\rho(r)}
\rlap{~~.}
\label{eq:rd}
\end{equation}
In Table \ref{table1} we give the value of $R_{2}$ and the density
$\rho(R_{2})$
calculated for various pion energies for $^{12}$C.  We note that over
this energy region (80~MeV $\le T_{\pi}\le$ 315~MeV) the interaction
is confined to the nuclear surface and low densities.  Here
$\rho_{0}=0.16$~fm$^{-3}$ (nuclear matter density) and the pion
distorted waves are taken from Ref. \cite{sici86}.

The Pauli exchange term can be taken directly from Ref. \cite{chiang85}
evaluated at the density $\rho(R_{2})$.  We extend the term by including
rho--meson propagation in the intermediate state.  We omit
pion distortions for the intermediate pion to avoid including
multiple reflection corrections in the Pauli term.
The $\lambda_{0}^{(2)}$ coefficients are
given in Table \ref{table1}.  The dotted curve in Fig. \ref{fig1} gives
differential cross sections resulting from adding the second--order Pauli
correction to the lowest--order calculation.  We see that the Pauli correction
is (a)~large and
(b)~completely destroys the nearly--quantitative agreement of
the dashed curve.

We will include true absorption by introducing a $\lambda_{0}^{(2)}$ parameter
determined from the spreading potential
of the delta--hole model.  These two terms cannot be equated directly because
the the spreading potential occurs in
the denominator of the delta propagator.  We can make the
correspondence by first isolating the $P_{33}$ partial--wave
contribution to the lowest--order optical potential and expressing it
in a resonant form.  The difference
between this potential evaluated twice,
once with width $\Gamma_{0}+Im\,W_{sp}$ and then with width
$\Gamma_{0}$ (the free width), is a true--absorption potential that can be
expanded at low density to give a $\lambda_{0}^{(2)}$
independent of $r$.  Rather than expanding, however, we determine
$\lambda_{0}^{(2)}$ by matching this difference to Eq. \ref{eq:nd}
at the radius $R_{2}$. The resulting
values of $\lambda_{0}^{(2)}$ are given in Table \ref{table1}.  We see
that, at all energies, there is a large cancellation between
$\lambda_{0}^{(2)}$ (Pauli) and $\lambda_{0}^{(2)}$ (spreading),
yielding a small total second--order correction.  The solid curve in Fig.
\ref{fig1} gives the differential cross sections obtained when
$E_{ms}$, $\lambda_{0}^{(2)}$ (Pauli), and $\lambda_{0}^{(2)}$
(spreading) are all included.  The cancellation
of the Pauli and spreading terms is evident.

Finally, we also include the correlation (or LLEE) corrections.
It has been shown \cite{johnson88} that the LLEE effect can be included in
the delta self--energy by a modification
\begin{equation}
\delta E_{ms}={4\xi\over 27}\left({f_{\pi N\Delta}\over m_{\pi}}
\right)^{2}\rho_{0}\rlap{~~,}
\label{eq:llee}
\end{equation}
where $\xi$ is the usual Lorentz--Lorenz parameter.  The value of $\xi$
depends on the range of the short--range repulsive correlations between
nucleons, the range of the pion--nucleon form factor, and the strength
of the delta--nucleon interaction.  We will allow for some uncertainty
in the LLEE--parameter $\xi$.  The minimum value that is reasonable is about
$\xi/3=0.08$, which results from a pion--nucleon monopole cutoff of
800~MeV/$c$ and no $\Delta N$ interaction.  The maximum value of
$\xi/3$ is 0.23, which arises from a cutoff of 990~MeV/$c$ and includes
a $\Delta N$--interaction contribution.  This is a value that would
give the real part of the delta--hole spreading interaction, which
corresponds to $\delta E_{ms}=23$~MeV.
The final result of this work is given by the shaded
area between the solid curves in Fig. \ref{fig2} (corresponding to
the range $0.08\le\xi/3\le 23$).  These results combine the
first--order potential in which the delta--nucleus potential is included
via the mean spectral energy with Pauli, true--absorption, and correlation
corrections.

The agreement with
the data shown in Fig. \ref{fig2} is not exact, but it is remarkably good
for
a parameter--free calculation.  Discrepancies could be due to the fact that our
treatment of the second--order
corrections is neither exact nor totally consistent (we have taken the
true--absorption term from the delta--hole model).  For these reasons,
it is probably unwarranted to conclude that the smaller value of
$\xi/3=0.08$ is preferred, even though this result is everywhere
closer to the data.  Firm conclusions should await a more thorough, internally
consistent treatment \cite{ann} of all the higher--order terms.  We are
motivated to pursue this treatment because our present calculation is
intriguingly close to the data.

We have for the first time combined a contemporary momentum--space
calculation of the first--order optical potential with microscopic
predictions of the effects of the delta--nucleus interaction, Pauli
corrections, pion true absorption, and short--range correlations.
We have seen that convergence of the expansions appears to be enhanced
throughout the resonance region by (1) collecting $U_{\Delta}$ (via the
mean spectral energy approximation)
together with binding corrections into the first--order
optical potential, and (2) collecting
the Pauli and true--absorption terms
together.  This result supports our perturbative approach \cite{ann} to
calculating the optical potential.
Finally, the good results that we find from 80 to 226~MeV with no
adjustable parameters suggest that pursuing calculations of
greater accuracy for the second--order terms might yield a
definitive determination of the short--range correlations (i.e.,
the parameter $\xi$) and the delta--nucleus interaction, $U_{\Delta}$.

\figure{The differential cross section for elastic scattering
of $\pi^+$ from $^{12}C$ at the energies indicated on the figure.
The dashed curves are a complete lowest--order optical--model calculation
including the delta--nucleus
interaction through the mean spectral approximation. Dotted:  the
complete lowest--order optical potential and the second--order Pauli
corrections are included; solid:  the complete lowest--order optical
potential, the second--order Pauli, and the second--order
spreading potential are all included.  The data are from Ref.
\cite{harvey84}.\label{fig1}}

\figure{The same as Fig. \ref{fig1}, except the shaded area
includes the full lowest--order optical
potential, Pauli and spreading corrections, and the LLEE correlation
corrections. The two curves forming the boundary result from the
LLEE parameter $\xi$
equal to .08 (the lowest curve in the forward direction) and equal to 0.23.
\label{fig2}}

\newpage
\mediumtext
\begin{table}
\caption{Parameters for the Pauli and spreading interaction.
The $\lambda^{(2)}_0$ as defined in Eq. \ref{eq:nd} is given as a function
of the pion kinetic energy $T_\pi$ (Mev) and corresponds to the the density
region centered about the radius $R_2$ (fm) in $^{12}C$. The units for
$\lambda^{(2)}_0$ are fm$^3$.}
\label{table1}
\begin{tabular}{rccccc}
\multicolumn{1}{c}{T$_{\pi}$} &\multicolumn{1}{c}{R$_2$}
&\multicolumn{1}{c}{$\rho~(R_2$)/$\rho_0$}
&\multicolumn{1}{c}{$\lambda^{(2)}_0$~(Pauli)}
&\multicolumn{1}{c}{$\lambda^{(2)}_0$~(spread)}
&\multicolumn{1}{c}{$\lambda^{(2)}_0$~(Sum)}\\
\tableline
 80&2.40&0.289&-0.40, -1.46&-0.93, ~2.02&-1.33, ~0.56\\
100&2.52&0.252&~0.08, -1.88&-1.43, ~2.09&-1.36, ~0.21\\
148&2.80&0.175&~2.50, -1.74&-3.29, ~1.18&-0.80, -0.56\\
162&2.86&0.160&~3.20, -0.90&-3.70, ~0.35&-0.50, -0.55\\
230&2.90&0.151&~0.54, ~2.50&-1.02, -2.80&-0.47, -0.31\\
315&2.67&0.209&-0.60, ~0.26&~0.67, -0.68&~0.07, -0.42\\
\end{tabular}
\end{table}
\end{document}